\documentclass[submission]{eptcs}
\providecommand{\event}{F-IDE 2019} % Name of the event you are submitting to
\usepackage{breakurl}             % Not needed if you use pdflatex only.
\usepackage{underscore}           % Only needed if you use pdflatex.
\usepackage{graphicx}
\usepackage{enumitem}
\usepackage{caption,subcaption}
\usepackage{tabularx}

\usepackage {oplotsymbl}

\title{An Integrated Development Environment for the Prototype Verification System}
\author{Paolo Masci
	\institute{National Institute of Aerospace\\Hampton, VA, USA}
	\email{paolo.masci@nianet.org}
	\and
	C\'esar A. Mu\~{n}oz
	\institute{NASA\\Hampton, VA, USA}
	\email{cesar.a.munoz@nasa.gov}
}
\def\titlerunning{An Integrated Development Environment for the Prototype Verification System}
\def\authorrunning{Paolo Masci \& C\'esar A. Mu\~{n}oz}

\begin{document}

\maketitle

\begin{abstract}
The steep learning curve of formal technologies is a well-known
barrier to the adoption of formal verification tools in industry. This
paper presents VSCode-PVS, a modern integrated development environment
for the Prototype Verification System (PVS). This new environment
integrates the editing and proof management functionalities of PVS in
Visual Studio Code, a popular code editor widely used by software
developers. VSCode-PVS provides functionalities that
developers expect to find in modern verification tools, but are not
available in the standard Emacs front-end of PVS, such as
auto-completion, point-and-click navigation of definitions, live
diagnostics for errors, and literate programming. The main features and
architecture of the environment are presented, along with a comparison
with other similar tools.
\end{abstract}
% Interactive Theorem Proving, Prototype Verification System,
% Integrated Development Environment

\section{Introduction}
Early detection of design anomalies and increased confidence that the
system will operate as intended are some of the benefits of the use of
formal verification technologies in industry.  However, outside
safety-critical domains such as avionics, the use of verification
tools is still rather limited. One of the reasons is the steep
learning curve of verification technologies, which creates an initial
cost that is often deemed excessive with respect to the long
term benefits.

The work presented in this paper aims to reduce the learning curve of the Prototype Verification System (PVS)~\cite{owre1992pvs},
a verification tool for formal modeling and analysis
of system designs. The PVS modeling language is based on higher-order
logic. It supports basic types (\texttt{boolean}, \texttt{int},
\texttt{real}, etc), as well as datatypes such as \texttt{string},
\texttt{set}, \texttt{list}. The proof engine is based on Gentzen's
sequent calculus, and supports the use of proof strategies for
automated analysis. The verification system provides an evaluation environment, called PVSio~\cite{munoz2003rapid},  for animation of
executable specifications. 
%Recently, a prototyping environment PVSio-web~\cite{masci2015pvsio} has also been developed that facilitates the development of simulated prototypes based on executable PVS specifications. 

PVS is a powerful analysis tool with a long history of success stories
in a range of different domains~\cite{munoz2012}. PVS is widely
used at  NASA Langley Research Center for the analysis of algorithms and protocols for avionics
systems\footnote{\url{https://shemesh.larc.nasa.gov/fm/fm-main-research.html}}. Research groups have also applied PVS to the analysis of
human-machine interfaces in medical systems~\cite{masci2014formal}
and for co-simulation of Cyber-Physical Systems~\cite{sosym19} and semi-autonomous
systems~\cite{palmieri2017co}.

Becoming fluent with PVS, however, usually requires several
weeks. Common difficulties faced by developers are often rooted in
the PVS front-end:
\begin{itemize}
\item {\bf Conceptual gap.} The PVS front-end is based on the Emacs
editor. Emacs %is different than other code editors typically used by
%developers. It 
does not provide separate visual components for editing
files, executing commands, and browsing the file system. Rather, it
provides {\em buffers}, abstract entities that can be used to interact
with any resource. When a buffer is linked to a text file, e.g., a PVS
specification, Emacs acts like a code editor. When the buffer is
linked to an interactive process, e.g., the PVS theorem prover, Emacs
provides a command-line interface for sending commands to the
process. Developers need to learn the set of commands for operating
with buffers, as well as to recognize the buffers --- the visual
appearance of different buffers is identical to the untrained eye.

\item {\bf Knowledge gap.} 
The PVS Emacs interface favors the use of
the command line integrated in Emacs. % to point-and-click interactions. 
Developers need to learn several commands and
keyboard shortcuts to be fluent when editing, parsing, and analyzing
PVS specifications. Means for efficient navigation of the %PVS
libraries are not provided in PVS Emacs. This can significantly slow
down the development of PVS specifications and proofs. For example,
the NASA PVS Library~\footnote{\url{https://shemesh.larc.nasa.gov/fm/ftp/larc/PVS-library}.} includes over 100,000
theorems and definitions useful for modeling and analysis of different
aspects of safety-critical systems. Without appropriate tool support,
finding anything within these libraries is a prohibitive task.
\end{itemize}

%\smallskip\noindent {\bf Contribution.~} 
The main contribution of this
work is VSCode-PVS, a new integrated development environment designed
to reduce substantially the conceptual and knowledge gaps faced by PVS
users. A preliminary but fully functional version of the new
environment is presented, along with a comparison with other
environments. The environment is publicly available at GitHub\footnote{\url{https://github.com/nasa/vscode-pvs}.} under
NASA's Open Source Agreement.

%\smallskip\noindent {\bf Organization.~} 
The rest of this paper is organized as follows. Section~\ref{sec:background} provides background information on the standard PVS front-end, and on Visual Studio Code. Sections~\ref{sec:vscode-pvs} and~\ref{sec:example}
introduce the architecture and main features of the environment.
Section~\ref{sec:comparison} identifies a set of metrics
that are general across the front-end of different modeling and
analysis tools, and uses these metrics to compare VSCode-PVS to other
environments. Section~\ref{sec:related} presents related
work. Finally, Section~\ref{sec:conclu} concludes the paper.

\section{Background}
\label{sec:background}
This section provides background information on PVS Emacs, the standard front-end of PVS, and Visual Studio Code, the baseline technology used for the development of VSCode-PVS.

\subsection{PVS Emacs}
PVS Emacs is a text-based environment built on the Emacs editor to support user interaction with PVS.
%The environment builds on the Emacs editor, which integrates word editing functionalities with an interactive command line, called minibuffer, for executing commands.
Commands necessary to interact with the PVS system are invoked by typing the command name and its arguments, if any, in the Emacs minibuffer.
To enter a command in the minibuffer, users need to pre-fix the command with the key sequence \texttt{M-x}, where \texttt{M} is usually the Alt key.
An example is \texttt{M-x typecheck}, which executes the PVS type-checker command on the PVS file opened in the Emacs editor. Over 100 commands are provided (see~\cite{pvs-system-guide}). Additionally, an extensive list of keyboard shortcuts is implemented to speed up command entry. For example, type-checking can be executed by typing
\texttt{M-x tc} in the minibuffer, or by performing the key sequence \texttt{C-c C-t}, where \texttt{C-} is the Ctrl key.
 
\subsection{Visual Studio Code}
Visual Studio Code is a cross-platform open-source code editor created in 2015 by Microsoft.
The editor provides a rich graphical user interface that integrates the essential components typically used by programmers:
a source code editor that supports auto-completion, hovers, embedded mini-editors, and contextual menus;
an integrated graphical debugger, which allows the user to set break-points and perform step-by-step execution of source code;
a tree-based view for browsing files in the file system;
an integrated terminal for executing commands; integration with source code management tools (e.g., Git).
The behavior of all these components can be extended or re-programmed
%with the aim 
to provide support for a specific programming language.
The community has already created extensions for over 80
programming languages, including C++, Java, JavaScript, and Python.
Besides programming languages, there is also a growing interest in
integrating verification tools in Visual Studio Code. Examples
include
Dafny\footnote{\url{https://github.com/DafnyVSCode/Dafny-VSCode}} and
Lean\footnote{\url{https://github.com/leanprover/vscode-lean}}.
%, and vscoq\footnote{\url{https://github.com/siegebell/vscoq}}. 
% and a Visual Studio Code extension for Isabelle\footnote{\url{https://github.com/seL4/isabelle/tree/master/src/Tools/VSCode/extension}}. 
Some of these extensions will be discussed further below, in
Section~\ref{sec:comparison}.

\section{VSCode-PVS}
\label{sec:vscode-pvs}
\begin{figure}[t]
	\centering
	\includegraphics[width=\linewidth]{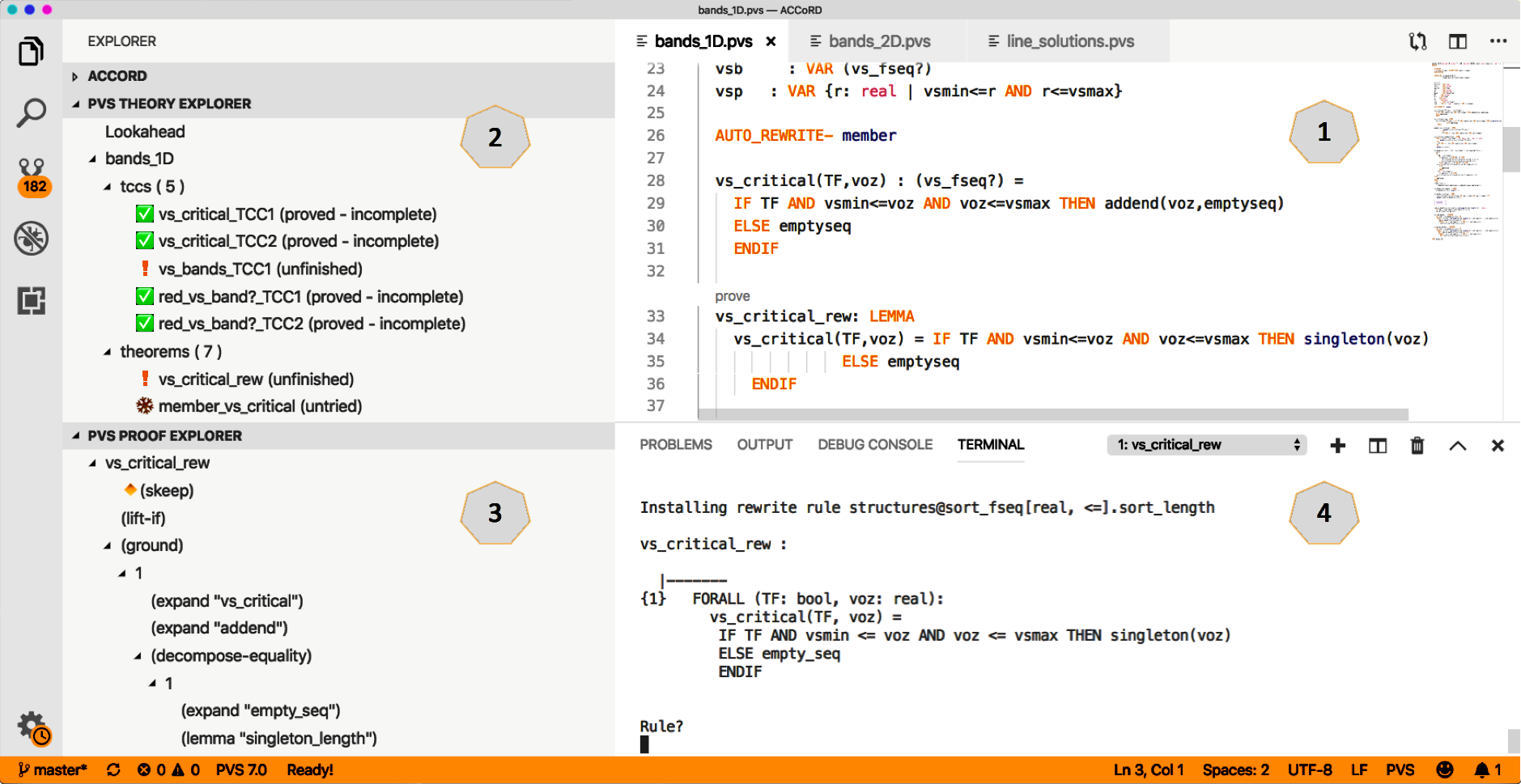}
	\caption{VSCode-PVS: (1) Main Editor; (2) Theory Explorer; (3) Proof Explorer; (4) Integrated PVS Command Line Interface (CLI).}
	\label{fig:vscode-pvs}
\end{figure}

VSCode-PVS is a new integrated development environment for creating, evaluating and verifying PVS specifications.
The environment, shown in Figure~\ref{fig:vscode-pvs}, redefines the way developers interact with PVS, and better aligns the PVS front-end to the functionalities provided by development environments used by software developers. The main features provided by the environment are as follows:

\begin{itemize}
	\item {\bf Syntax highlighting.} PVS keywords and library functions are automatically highlighted.
	\item {\bf Autocompletion and code snippets}. Tooltips suggesting function names and language keywords are automatically presented when the user types a symbol in the editor. Code snippets are provided for frequent modeling blocks, e.g., \texttt{if-then-else}. 
	\item {\bf Hover information for symbol definitions}. Hover boxes providing information about identifiers are automatically displayed when the developer places the cursor over an identifier.
	\item {\bf Jump-to declaration}. Navigation of symbol declarations can be performed with simple point-and-click actions: the user places the cursor over the name of an identifier, and a click on the name of the identifier while holding the Ctrl key down opens a window with the location where the identifier is declared.
	\item {\bf Live diagnostics}. Parsing is automatically performed in the background, and errors are reported in-line in the editor. Problematic expressions are underlined with red wavy lines. Tooltips presenting the error details are shown when the user places the cursor over the wavy lines. 
	\item {\bf In-line actionable commands}. Actionable commands are available for PVS theorems. They are rendered in-line in the editor, above the name of the theorem, and can be used to start a new prover session for the theorem with a simple click action.
	\item {\bf Overview of PVS theories}. The overall structure of a set of PVS theories is rendered using an interactive tree-based view. It shows the set of PVS theories in the active workspace, as well as the name and status (proved, unfinished, etc.) of the theorems defined in each theory. Point-and-click actions can be used to jump to theory definitions and type-check the theories.
	\item {\bf Interactive proof tree visualizer and editor}. An interactive tree-based view shows the proof associated with a theorem. Point-and-click actions are provided for step-by-step execution of proof commands. Functionalities for editing the proof are currently under development.
	\item {\bf Integrated PVS and PVSio Command Line Interfaces}. Integrated command line interfaces allow interaction with the theorem prover and the PVSio evaluator. Auto-completion is provided for prover commands, as well as access to the commands history.
\end{itemize}

\subsection{Architecture}
\begin{figure}[t]
	\centering
	\includegraphics[width=0.8\linewidth]{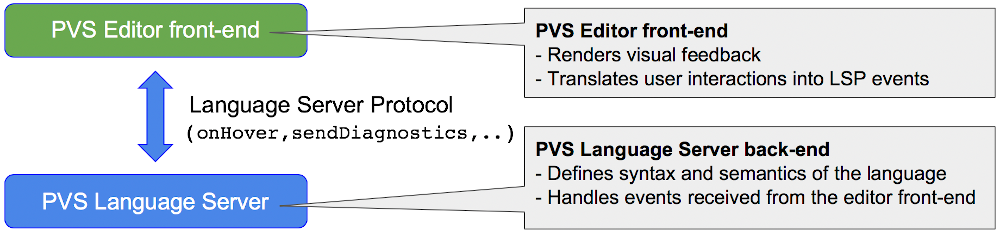}
	\caption{Overall architecture of VSCode-PVS.}
	\label{fig:architecture}
\end{figure}

The overall architecture of VSCode-PVS is depicted in Figure~\ref{fig:architecture}. It builds on the Language Server Protocol %\footnote{\url{https://langserver.org}} 
(LSP), a tool-independent communication protocol for exchanging data and events between two architectural elements: an editor front-end and a language server back-end.
 
The {\em editor front-end} is responsible for rendering visual feedback to the user, and transforms user interactions with the editor into corresponding LSP events to be dispatched to the language server. The {\em language server} defines the functions necessary to support the syntax and semantics of the language (e.g., in the case of PVS, parsing, typechecking, etc.) and continuously listens to LSP events. An example LSP event is \texttt{onHover}. This event is triggered by the editor front-end when the user places the cursor over an identifier in the text document. This event is sent to the language server, along with information on the path of the text document and the location of the cursor in the document. The language server acts upon this event, in this case, by sending the identifier's definition back to the editor. The editor, in turn, displays the server response to the user as a hover box. The language server can also generate events. For example, an event \texttt{sendDiagnostics}, is used by the server to publish diagnostics information (e.g., parsing error).

The LSP protocol is extensible. It builds on Remote Procedure Calls (RPCs) and the JavaScript Object Notation (JSON) format. It allows the definition of new event types to accommodate language-specific features. In the case of PVS, this feature is used to implement commands necessary for interactive analysis of PVS specifications, e.g., type-checking, discharging proof obligations, proving theorems, etc.
The LSP-based architecture has been chosen for the implementation of VSCode-PVS because it promotes reuse of modules and facilitates sustainability of the overall development effort --- LSP-compliant editors can be connected to the Language Server back-end. 
%The developers community is actively contributing to the LSP architecture, e.g., Language Servers are already available for 50+ programming languages. 
All major code editors support the LSP, which makes it relatively simple to connect the PVS language server to a different editor front-end. Features implemented in the language server for standard LSP events become automatically available in the connected editor.

\subsection{The Editor Front-End}
\begin{figure}[t]
	\centering
	\includegraphics[width=0.9\linewidth]{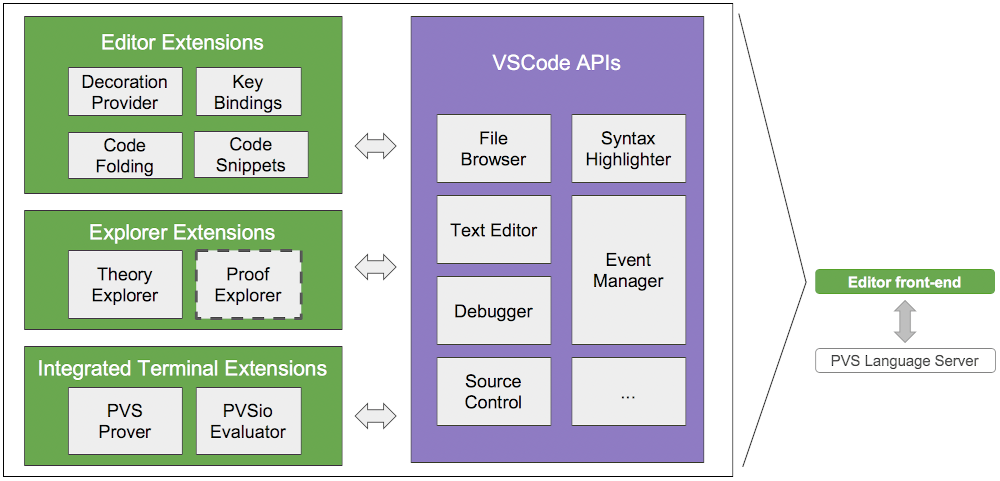}
	\caption{Inner architecture of the PVS editor front-end. Modules are represented as boxes. Communication between modules is indicated with arrows. Dashed line indicates under development.}
	\label{fig:architecture-client}
\end{figure}

The PVS Editor front-end builds on Visual Studio Code\footnote{\url{https://code.visualstudio.com}}, an open-source code editor widely used by software developers.
The main software modules of the front-end are shown in Figure~\ref{fig:architecture-client}. They are illustrated in the following.

\smallskip\noindent
{\bf Editor Extensions.} These modules are used to tailor the functionalities of the textual editor provided by Visual Studio Code to the PVS language. Functionalities implemented in these modules include: decorators necessary for syntax highlighting; code snippets for rapid creation of PVS code blocks; code folding; key bindings and contextual menus for PVS commands.

\smallskip\noindent
{\bf Explorer Extensions.} These modules customize Explorer View, a graphical tree-based view integrated in Visual Studio Code. A new module, {\em Theory Explorer}, introduces support for click-and-point navigation of PVS specifications.
%The root node in the tree view provided by Theory Explorer is the PVS context folder, i.e., the directory containing the PVS files being developed or analyzed in the current session. The first level of nodes in the tree identify the set of PVS theories loaded in the PVS systems. The second level of nodes represent proof obligations. Point-and-click operations can be used to jump to a theory definition within a file. In-line commands and contextual menus provide access to common functionalities such as parsing, type-checking, proving. 
A second module, {\em Proof Explorer}, aims to support interactive visualization and editing of proofs using click-and-point operations.

\smallskip\noindent
{\bf Integrated PVS Terminals.} These modules seamlessly link the integrated terminal of Visual Studio Code to the interactive read-eval-print loops of the PVS theorem prover and the PVSio~\cite{munoz2003rapid} evaluator. That is, in the {\em prover terminal}, developers can type proof commands for the theorem prover, and watch the proof state returned by the prover directly in the terminal. Similarly, in the {\em evaluator terminal}, developers can type ground expressions and thus execute fragments of a PVS specification.

\smallskip\noindent
{\bf VSCode APIs.} This is a library provided by Visual Studio Code for extending and customizing the functionalities of the editor. It includes communication primitives necessary to support the LSP.

\subsection{The Language Server}
\begin{figure}[t]
	\centering
	\includegraphics[width=0.84\linewidth]{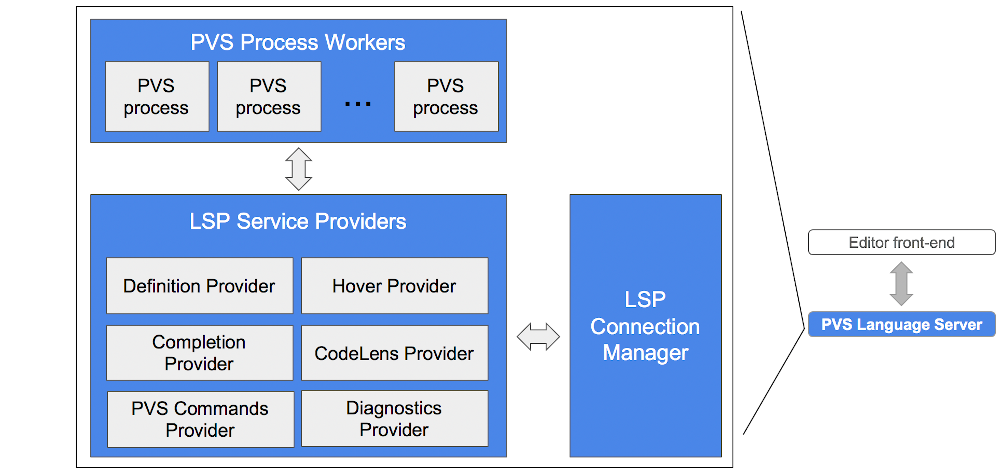}
	\caption{Inner architecture of the PVS Language Server.}
	\label{fig:architecture-server}
\end{figure}

The PVS Language Server includes the following modules (see Figure~\ref{fig:architecture-server}):

\smallskip\noindent
{\bf LSP Service Providers.} These modules handle LSP events received by the language server. They reconcile the APIs of PVS with the logic of the LSP. For example, 
{\em Hover Provider}, which is activated when an \texttt{onHover} event is received, uses the APIs of PVS to gather the information to be shown in hover boxes, and then sends the response to the editor front-end using the LSP format.
{\em Definition Provider} defines the logic necessary to support functions such as go-to definition and peek definition, which are used in the editor front-end for point-and-click navigation of PVS specifications.
{\em CodeLens Provider} defines the logic behind actionable commands embedded in-line in the PVS specification --- this is used to introduce in-line actionable {\em prove} commands at the location of theorem definitions.
{\em PVS Commands Provider} provides support for language-specific commands used for analysis of PVS specifications, e.g., type-check, show proof obligations, prove theorem.
{\em Diagnostics Provider} is a background process that continuously sends diagnostics information to the editor front-end to report syntax / type-checking errors.

\smallskip\noindent
{\bf LSP Connection Manager.} This is a routing module for managing the exchange of events and data with the client front-end. The module listens for client connections, receives LSP events from connected clients, and dispatches the events to the appropriate service provider.

\smallskip\noindent
{\bf PVS Process Workers.} These modules embed the functionalities of the PVS verification system in the language server. Each worker executes a PVS instance in a self-contained execution environment. A pool of workers supports parallel execution of multiple PVS instances --- this is used for running background services like parsing and type-checking. Furthermore, in contrast to PVS Emacs, VSCode-PVS  supports running simultaneously proofs of different formulas.

\subsection{Implementation}
VSCode-PVS is implemented in TypeScript\footnote{\url{https://www.typescriptlang.org}}, an extension of JavaScript that supports type annotations. Typescript programs can be statically checked for type correctness. A transpiler translates TypeScript code into plain JavaScript, which allows the execution of TypeScript programs in standard JavaScript engines. The editor front-end builds on the APIs of Visual Studio Code. The language server builds on NodeJS\footnote{\url{https://nodejs.org}}, a JavaScript environment that provides libraries necessary for creating web-services, including functions for spawning processes and performing operations on file systems. Process workers use the native Lisp interface of PVS to exchange commands and data with the PVS system. TSLint\footnote{\url{https://palantir.github.io/tslint}}, a static analyzer for TypeScript, is routinely used for checking compliance with established coding conventions. Jasmine\footnote{\url{https://jasmine.github.io}} is used for testing the APIs of the developed modules.

The implementation effort to date amounts to 7K LoC (3K LoC for the editor front-end, and 4K LoC for the language server). Most of the code developed for the editor front-end is associated with the interactive tree-based views for visualizing theories and proofs. The server back-end required the development of a Lisp interface to dialogue with PVS, as well as additional logic for LSP events that are not directly supported by PVS, e.g., auto-completion for identifiers, hover information, live diagnostics.

\section{Example use of VSCode-PVS}
\label{sec:example}
This section showcases the main features of VSCode-PVS for two representative tasks typically carried out by PVS users.
Tasks are provided in the form of short descriptions presenting the overall {\em goal} of the task and the {\em context} within which the task is carried out.
A comparison with the standard PVS Emacs front-end is included in each task to better appreciate the improvements introduced by VSCode-PVS.

\subsection{Task 1: Navigation of symbol declarations}
\noindent
{\bf Goal:} Inspect function and type declarations from imported PVS files.

\smallskip\noindent
{\bf Context:} When developing a new PVS specification or a new proof, users typically need to navigate symbol declarations imported from other PVS theories. This is necessary, e.g., to inspect the structure of complex datatypes defined in the PVS libraries.

\smallskip\noindent
{\bf Workflow with VSCode-PVS}. The standard workflow for inspecting a symbol declaration involves using the hover functionality (see Figure~\ref{fig:hover}). The user can place the cursor over the symbol definition, and a tooltip will be automatically shown. The tooltip includes three main elements: a brief description of the symbol (e.g., built-in type); a clickable link for jumping to the location of the declaration; a preview of the symbol declaration. This standard workflow can always be adopted when a PVS specification type-checks correctly. An alternative workflow is also available for theories that are still not type-checked, e.g., because the user has not finished yet typing the content of the specification. In these cases, the resolution of symbol declarations can still be performed, but can be less accurate when symbol names are overloaded. That is, when the symbol to be resolved is overloaded and the theory is not type-checked, an array of candidate declarations is presented. In these cases, the {\em peek declaration} functionality is employed (see Figure~\ref{fig:peek}). It opens a mini-editor and a file browser in the current editor window that can be used to inspect the candidate declarations.

\smallskip\noindent
{\bf Workflow with Emacs}. Symbol declarations can be inspected with the command \texttt{show-declaration}. The command takes the name of the symbol as argument. The current location of the cursor can be used to auto-complete the symbol name. The command opens a new Emacs buffer with a preview of the declaration. The command \texttt{goto-declaration} can then be used to jump to the location of the declaration. The command takes the name of the symbol as argument. When the theory is not type-checked, a command \texttt{find-declaration} can be used to inspect a list of possible candidates. The jump-to functionality is however not available in this case, and the user needs to manually open the file (command \texttt{C-c C-f} followed by the filename) and scroll the text to the position of the declaration.

\subsection{Task 2: Proving a theorem}
\noindent
{\bf Goal:} Verify that a PVS specification satisfies given formal (mathematical) properties.

\smallskip\noindent
{\bf Context:} For complex systems, it is important to analyze a system design before the actual system is built. This helps developers gain confidence that the system design complies to given specifications, and identify and fix potential design issues early in the development process, when the cost of design changes is still relatively low. In safety-critical application domains, such as avionics and healthcare, such design analysis is usually mandated by regulatory frameworks. Proving mathematical theorems that capture properties of the intended characteristics and functionalities of the system is the core approach used in formal methods. It provides means to check properties of a system design for all possible inputs in all possible system states. 

\smallskip\noindent
{\bf Workflow with VSCode-PVS}. In-line actionable commands are provided next to each theorem. For example, in Figure~\ref{fig:vscode-pvs}, an actionable command \texttt{prove} is shown in the main editor window, above the theorem name at line 33). A click on the actionable command, triggers type-checking, and launches a new prover sessions in the integrated PVS Command Line Interface (see Figure~\ref{fig:vscode-pvs}, lower-right panel). Proof commands can be typed in the command line interface. The current proof is shown on the side, using an interactive tree-based view, called Proof Explorer (see Figure~\ref{fig:vscode-pvs}, lower-left panel). Nodes in the tree view can be collapsed/expanded to facilitate inspection of large proofs. Editing of the proof tree from Proof Explorer is under development.

\smallskip\noindent
{\bf Workflow with Emacs}. The user needs to type-check the file (\texttt{M-x tc}) and then start a theorem prover session with the command \texttt{M-x prove}. The command opens a new Emacs buffer that can be used to interact with the theorem prover. Proof commands are typed in this new buffer. A command \texttt{M-x x-show-current-proof} can be used to start open a window showing the proof tree (see Figure~\ref{fig:emacs}). The proof tree cannot be edited, and it does not allow collapsing/expanding of proof branches.

\section{Comparing VSCode-PVS to Other Analysis Tools}
\label{sec:comparison}
This section presents a comparison between VSCode-PVS and other similar environments. The following verification environments are considered:
\begin{itemize}[noitemsep]
\item PVS Emacs, the standard front-end of PVS;
\item Isabelle/jEdit~\cite{wenzel2012isabelle,wenzel2018isabelle}, the standard front-end of the Isabelle/HOL theorem proving system;
\item SublimeHOL\footnote{\url{https://github.com/JamesShaker/SublimeHOL}}, a front-end to the HOL4 theorem prover;
\item CoqIDE, the default front-end of the Coq theorem proving systems;
\item Proof General~\cite{aspinall2000proof}, a generic front-end for theorem provers;
\item Lean~\cite{de2015lean}, a new open-source theorem proving system from Microsoft;
\item IntelliJ-Arend\footnote{\url{https://github.com/JetBrains/intellij-arend}}, a new proof assistant under development at JetBrains research;
\item KeYmaera-X~\cite{fulton2015keymaera}, an interactive theorem prover for hybrid systems.
\end{itemize}

\subsection{Metrics}
A set of metrics has been defined to guide the comparison. The set of metrics is not exhaustive. 
Rather, they capture core functionalities necessary to support common modeling and analysis tasks.
\begin{enumerate}[noitemsep]
\item {\bf Autocompletion}: ability to suggest keywords and identifiers while typing; 
\item {\bf Hover information}: ability to show informative pop-ups for keywords and identifiers; 
\item {\bf Jump-to-definition}: ability to open a file at the location of an identifier's definition;
\item {\bf Refactoring}: ability to rename identifiers;
\item {\bf Live diagnostics}: ability to show diagnostics information while typing;
\item {\bf Animation}: ability to evaluate executable specifications;
\item {\bf Proof visualizer}: ability to visualize a proof tree.
\end{enumerate}

\begin{figure}[t]
\begin{subfigure}[b]{0.5\textwidth}
   \centering 
   \includegraphics[trim={0 0 6cm 0},clip,scale=0.5]{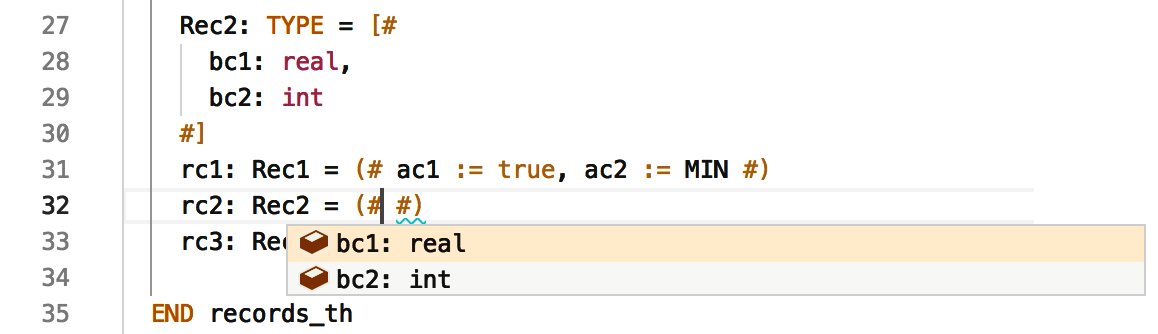}
   \caption{Autocompletion.} 
   \label{fig:autocompletion}
\end{subfigure}% 
\begin{subfigure}[b]{0.5\textwidth}
   \centering 
   \includegraphics[scale=0.5]{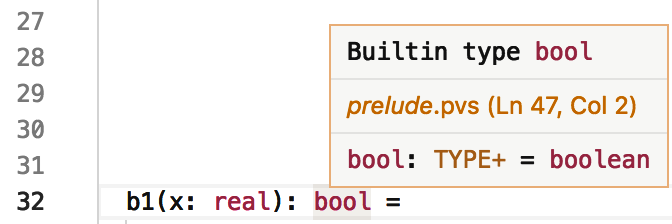}
   \caption{Hover.}
   \label{fig:hover}
\end{subfigure}%
\vspace{20pt}
\begin{subfigure}[b]{\textwidth}
   \centering 
   \includegraphics[scale=0.5]{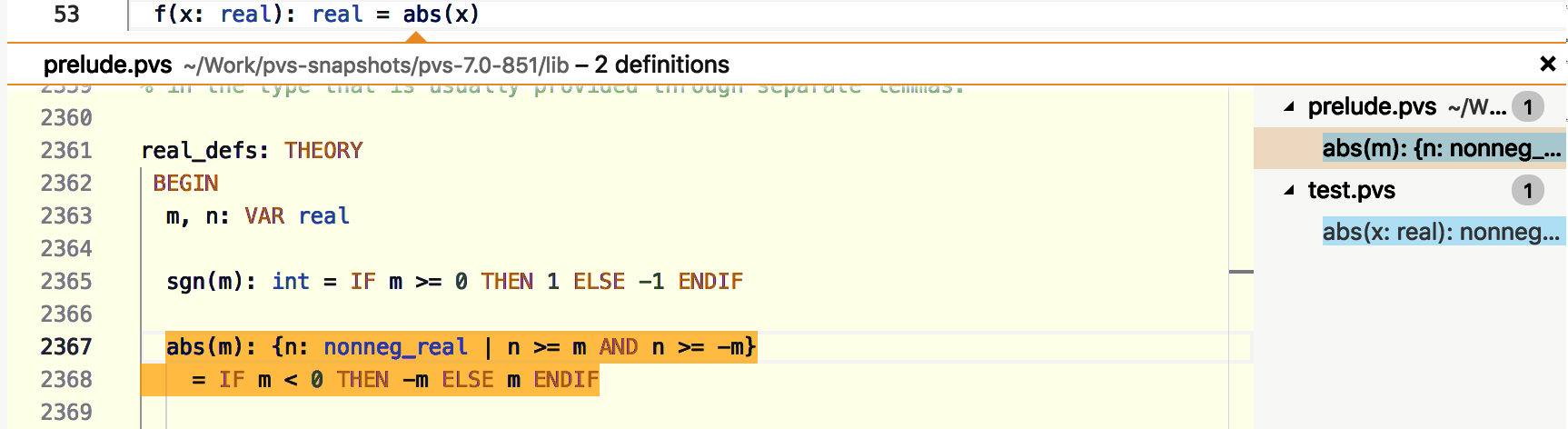}
   \caption{Peek definitions.}
   \label{fig:peek}
\end{subfigure}%
   \caption{Example functionalities of VSCode-PVS.} 
\end{figure}

%\begin{figure}[t]
%\begin{subfigure}[b]{0.6\textwidth}
%   \centering 
%   \includegraphics[scale=0.41]{figures/emacs-show-declaration}
%   \caption{Show declaration.} 
%   \label{fig:emacs-show-declaration}
%\end{subfigure}% 
%\begin{subfigure}[b]{0.4\textwidth}
%   \centering 
%   \includegraphics[scale=0.3]{figures/emacs-proof-viz}
%   \caption{Proof visualizer.}
%   \label{fig:emacs-proof-viz}
%\end{subfigure}%
%   \caption{Example functionalities of PVS Emacs.} 
%\end{figure}

%\begin{figure}[t]
%   \centering 
%   \includegraphics[trim={0.1cm 0.6cm 0 0.1cm},clip,width=0.9\textwidth]{figures/emacs}
%         \caption{PVS Emacs: editor and tree-based view.}
%   \label{fig:emacs} 
%\end{figure}

%\begin{figure}[t]
%   \centering 
%   \includegraphics[scale=0.5]{figures/coq}
%         \caption{CoqIDE.}
%   \label{fig:coq} 
%\end{figure}
%
%\begin{figure}[h]
%   \centering 
%   \includegraphics[scale=0.3]{figures/isabelle}
%         \caption{Isabelle/jEdit.} 
%   \label{fig:isabelle}
%\end{figure}

\begin{figure}[t]
\begin{subfigure}[b]{0.472\textwidth}
   \centering 
   \includegraphics[trim={0 0 0 4cm},clip,width=\textwidth]{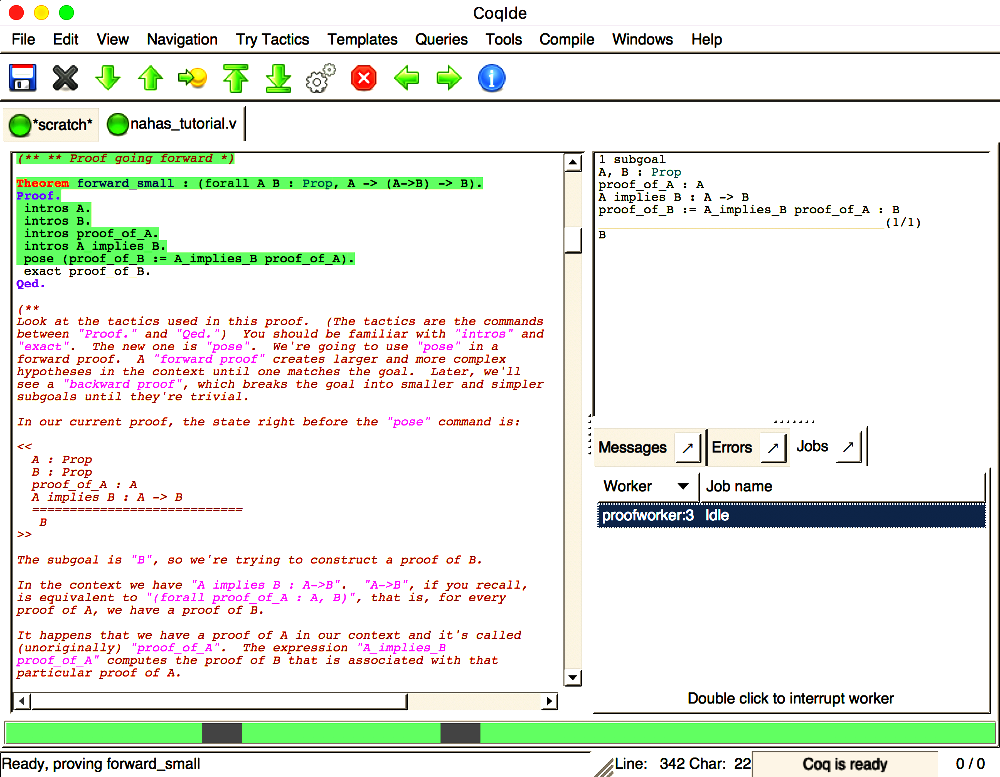}
   \caption{CoqIDE: editor and side view for proof inspection.} 
   \label{fig:coq}
\end{subfigure}% 
\hspace*{10pt}
\begin{subfigure}[b]{0.5\textwidth}
   \centering 
   \includegraphics[trim={0.06cm 0.1cm 0.05cm 1.7cm},clip,width=\textwidth]{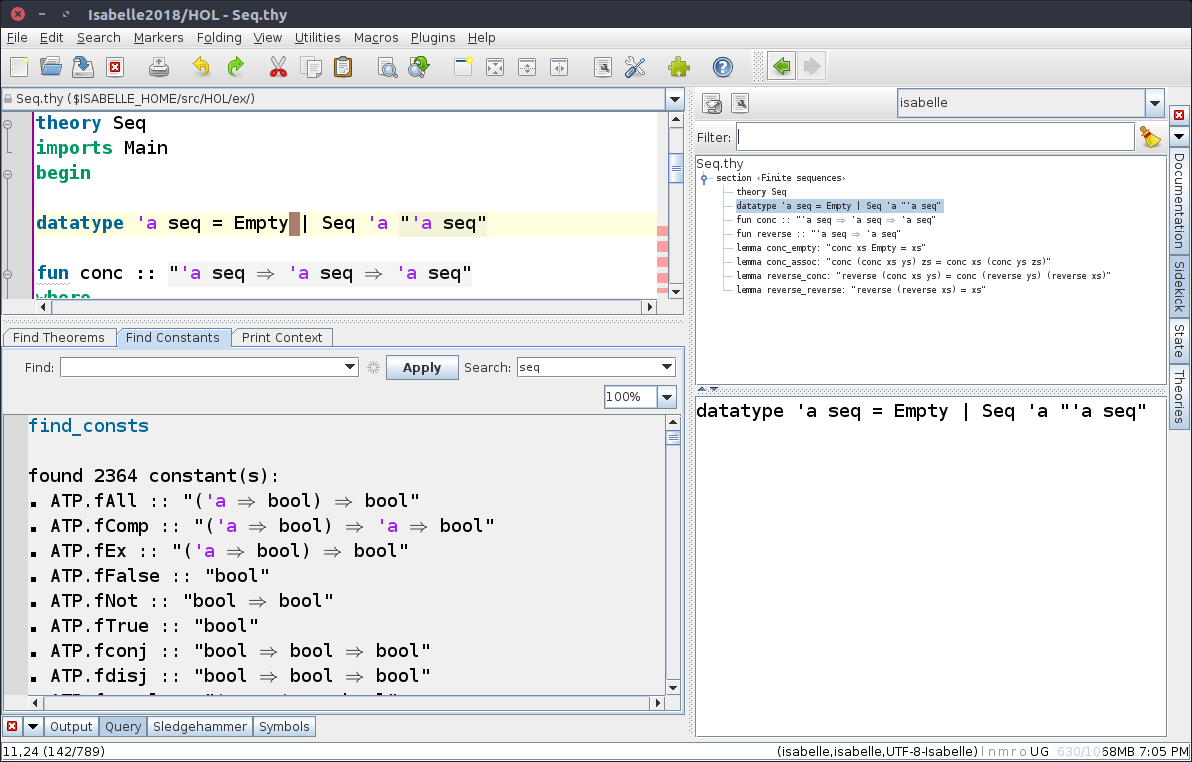}
   \caption{Isabelle/jEdit.}
   \label{fig:isabelle}
\end{subfigure}%
~\\~\\
\begin{subfigure}[b]{0.51\textwidth}
   \centering 
   \includegraphics[width=\textwidth]{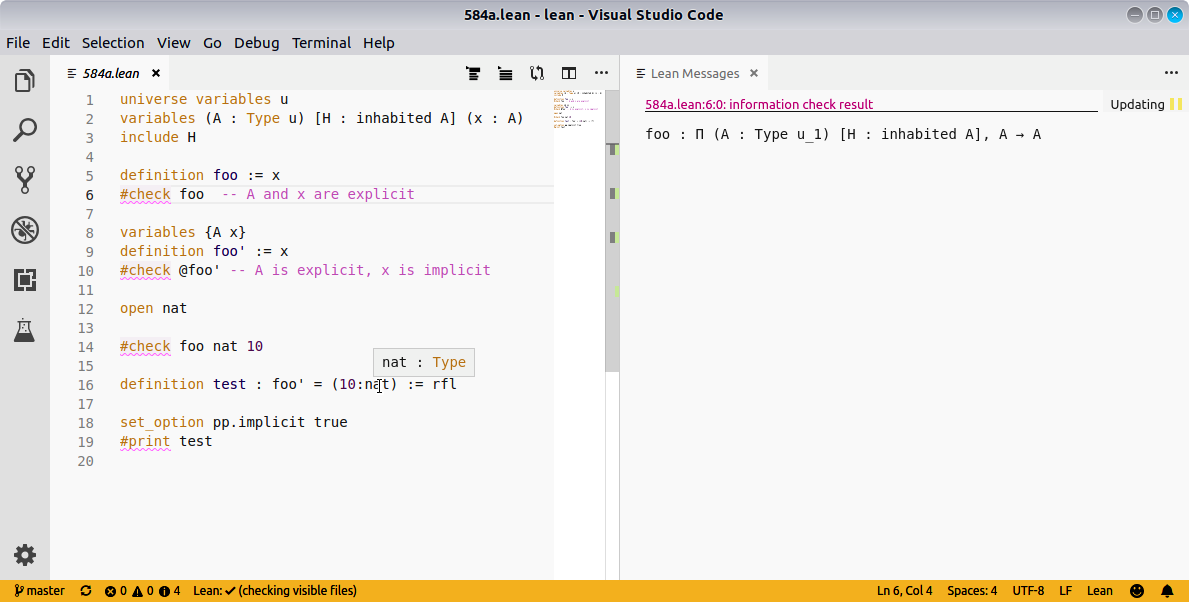}
   \caption{Lean.}
   \label{fig:lean}
\end{subfigure}%
\hspace*{10pt}
\begin{subfigure}[b]{0.46\textwidth}
   \centering 
   \includegraphics[width=\textwidth]{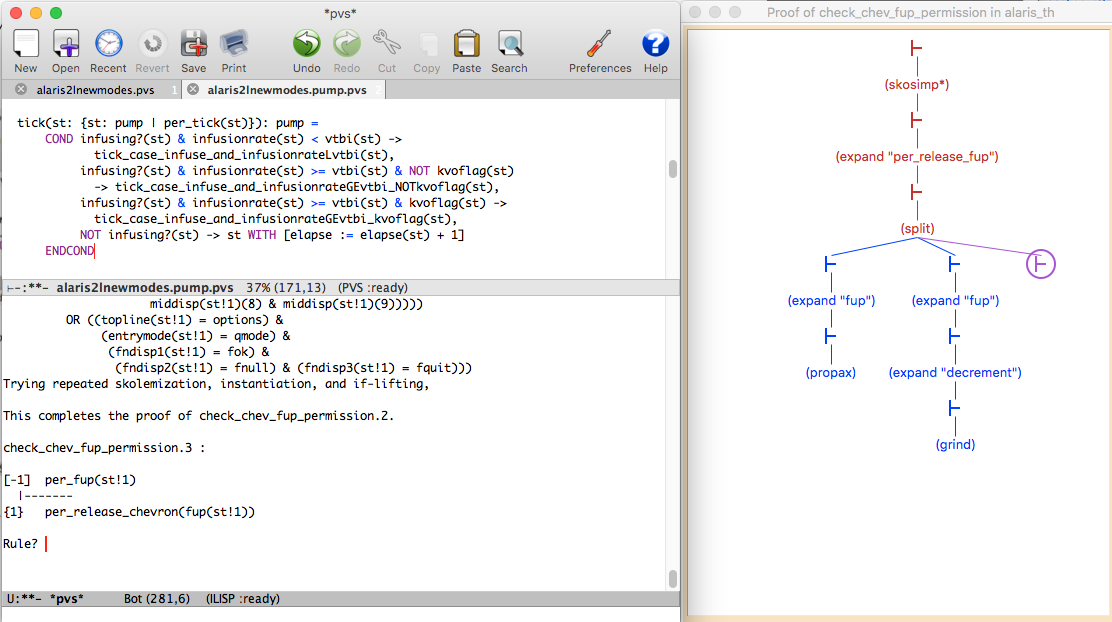}
   \caption{PVS Emacs: editor and proof window.}
   \label{fig:emacs}
\end{subfigure}%
~\\~\\
\begin{subfigure}[b]{0.51\textwidth}
   \centering 
   \includegraphics[width=\textwidth]{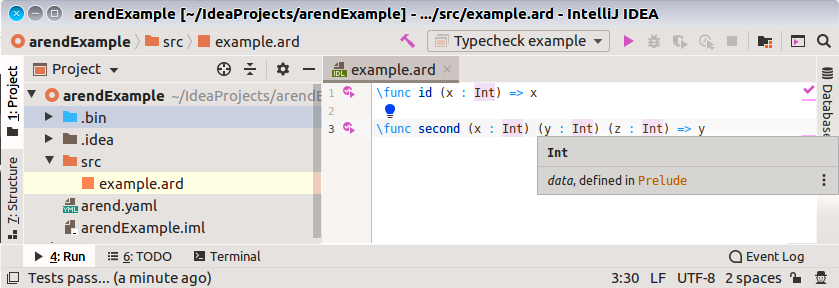}
   \caption{IntelliJ-Arend.}
   \label{fig:arend}
\end{subfigure}%
\hspace*{10pt}
\begin{subfigure}[b]{0.46\textwidth}
   \centering 
   \includegraphics[width=\textwidth]{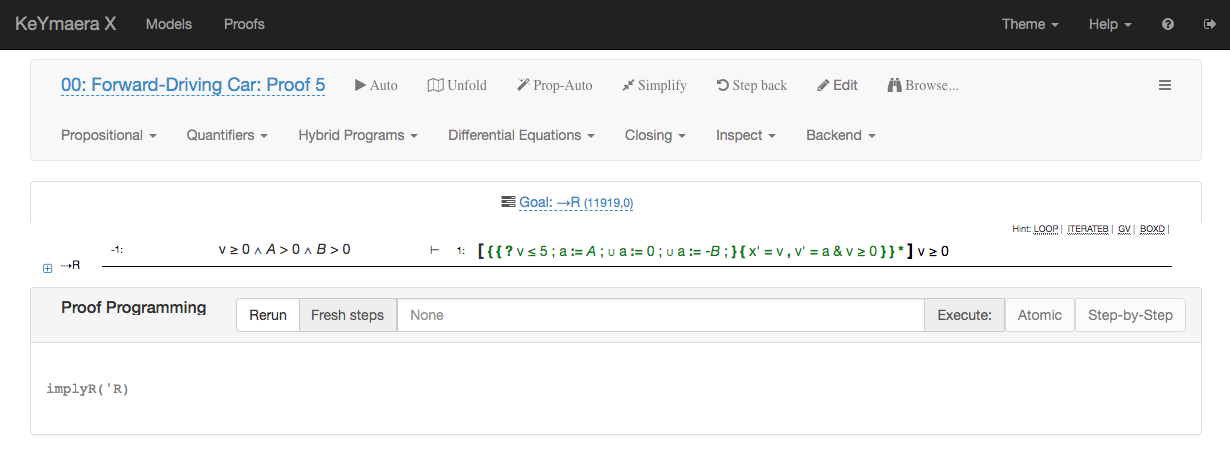}
   \caption{KeYmaera-X.}
   \label{fig:keymaerax}
\end{subfigure}%

   \caption{Screenshots of other analysis tools.} 
\end{figure}

%\begin{figure}[t]
%   \centering 
%   \includegraphics[scale=0.3]{figures/lean}
%   \caption{Lean.}
%   \label{fig:lean}
%\end{figure}
\clearpage

\subsection{Assessment}
{\bf VSCode-PVS}. 
Autocompletion is provided for language keywords, as well as for types and functions defined in the standard PVS library (the \texttt{prelude}).
Context-sensitive autocompletion is supported for record types: items suggested by the editor range over record accessors (see Figure~\ref{fig:autocompletion}).
Hover information is provided for identifiers. It shows a preview of the definition of the identifier, as well as an hyperlink that can be used to jump-to the location of the definition (see Figure~\ref{fig:hover}).
Live diagnostics for syntax errors are automatically provided after a period of inactivity in the editor --- syntax errors are underlined with red wavy lines, and a tooltip with details on the error is provided when placing the cursor at the error position. Integrated terminals can be used to animate specifications and prove theorems. An interactive tree-based view allows for the visualization of proof trees and proof commands step-by-step.

\smallskip\noindent
{\bf PVS Emacs}. The environment, shown in Figure~\ref{fig:emacs}, provides syntax highlighting and a basic form of jump-to-definition through commands and key bindings. A basic form of autocompletion is provided for proof commands. The logic for hover information is not implemented. Diagnostics are obtained on-demand, when the developer decides to parse or type-check the specification. Animation is provided through an interactive read-eval-print loop. Proofs can be visualized in a tree-based view. Nodes in the tree represent either proof sequents or proof commands. An interactive text-based view is used to show proof tactics and allows the user to perform step-by-step execution of proof commands.

\smallskip\noindent
{\bf CoqIDE}. The environment is shown in Figure~\ref{fig:coq}. It provides syntax highlighting. Autocompletion, hover information, and jump-to-definition are not provided. Diagnostics are provided on-demand, when the developer decides to attempt verification. A linear workflow is enforced: verification always starts from the beginning of the file, and a marker is advanced in the editor to indicate what has been verified. Everything above the marker is locked and cannot be modified. A side panel shows the proof state, including current goal, tactic state, and error messages. Point-and-click operations allow step-by-step execution of proof commands. The development of an alternative front-end, vscoq\footnote{\url{https://github.com/siegebell/vscoq}}, based on Visual Studio Code was attempted in recent years. However, the implementation appear to have stopped at the very early stages and does not provide significant new functionalities with respect to CoqIDE. 

\smallskip\noindent
{\bf Proof General}. The environment builds on Emacs, and its visual appearance is similar to PVS Emacs. The main design goal of this environment is to facilitate the execution of proof commands and the navigation of proof strategies. A plug-in based architecture allows the introduction of language-specific extensions. Plug-ins for Coq and Isabelle/HOL are available. They provide syntax highlighting and basic forms of autocompletion. Proof visualization builds on a tree-based view similar to that used in PVS Emacs. Commands are provided for step-by-step execution of proof tactics.

\smallskip\noindent
{\bf SublimeHOL}. The environment builds on the Sublime\footnote{\url{https://www.sublimetext.com}} editor. It provides syntax highlighting for the HOL4 specification language. A basic form of autocompletion is provided for keywords and mathematical symbols. Hover information, jump-to-definition and visualization of proof trees are not provided. Interactive panels allow to exchange commands with HOL4 and edit/inspect the proof state. % using a read-eval-print loop.

\smallskip\noindent
{\bf Isabelle/jEdit}. The environment, shown in Figure~\ref{fig:isabelle} builds on jEdit\footnote{\url{http://www.jedit.org/}}. Context-sensitive autocompletion is provided based on the syntax of the language, as well as on name-space information provided by the prover engine. A dictionary-based spell-checker is used to suggest completion items for comments and other sections in the specification that contain sentences in natural language.
Hover information shows the type of language symbols. Hover boxes can be detached from the current editor and turned into separate windows to facilitate navigation of file content.
Live diagnostics are provided for errors in the form of red wavy lines, along with hover information at the error location. A panel shows proof information and allows to perform step-by-step evaluation of a proof method. A query panel enables the filtering of information displayed for the proof state. A basic form of model animation can be achieved through Isabelle's counter-example finding functionality.

\smallskip\noindent
{\bf Lean}. The environment is shown in Figure~\ref{fig:lean}. It builds on Visual Studio Code. Because of this, its overall visual appearance is similar to VSCode-PVS. Context-sensitive autocompletion is provided for language symbols. A reasoning engine is always active in the background and tries to autocomplete expressions based on context information, in a manner similar to the {\em hole} functionality provided in Haskell. Hover information shows the definition of identifiers. Navigation of definitions is supported by the jump-to functionality and mini-editors rendered in line in the current editor. Live diagnostics are provided for errors, in the usual form of red wavy lines at the location of the error, and a tooltip with information about the error. Proof commands are embedded in the specification and can be activated with point-and-click operations. Similarly to Coq and jEdit, the proof state is shown in a side panel.

\smallskip\noindent
{\bf IntelliJ-Arend}. The environment builds on IntelliJ IDEA\footnote{\url{https://www.jetbrains.com/idea/}}. Autocompletion is provided for keywords and identifiers. An {\em auto import} function is available that automatically imports library modules. Hover information is used to present a description in natural language of identifiers, as well as a link to the definition of the identifier (see Figure~\ref{fig:arend}). Refactoring allows to rename identifiers, and move definitions across modules. Live diagnostics are presented for syntax errors. Additional diagnostics can be obtained on-demand by type-checking the specification. Animation and proof visualization are not available.

\smallskip\noindent
{\bf KeYmaera-X}. The environment is the successor of the KeYmaera IDE~\cite{platzer2008keymaera}. Web-based technologies are used to implement the front-end (see Figure~\ref{fig:keymaerax}). The main focus is on proof development. Only basic functionalities are provided to support modeling tasks: syntax highlighting is provided only for language keywords; hover information, jump-to definition and refactoring are not available. The prover interface renders proofs in sequent form, with horizontal lines as in the Gentzen-style layout. An extensive set of menus provides access to all proof commands. Heuristics are used to suggest proof tactics. % during proof attempts.

\subsection{Results}
An overview of the comparison is in Table~\ref{tab:results}. It can be seen that VSCode-PVS already provides several features that other similar environment are still missing.
Most of the environments are mainly designed to provide an interface for exchanging proof commands with the theorem prover. Few environments provide adequate support for modeling activities.
For example, almost all environments currently lack {\em refactoring}, and developers need to rely on search-and-replace functionalities of the editor when renaming identifiers. However, this solution is not robust, as careful inspection is necessary for overloaded identifiers.
Two other important features commonly used during modeling activities are also missing in most of the environments: {\em animation} of executable fragments and {\em live diagnostics}. 
Animation provides a means to developers to test a specification, e.g., to check whether it correctly captures what the developer wants to model. This functionality can be especially useful to software engineers that approach formal verification, as it is congruent with the testing methods they routinely use for software. It also provides a form a lightweight formal verification --- properties can be checked for specific execution traces prior to running the full formal proof.
Live diagnostics promote immediate identification of specification errors. This may facilitate understanding and resolution of errors, as a developer's focus of attention is already at the location of the error.

\begin{table}
\centering
\footnotesize
\begin{tabularx}{\textwidth}{rccXXXXXXXX}
 & \rotatebox{90}{Version} & \rotatebox{90}{Base editor} & \rotatebox{90}{Autocompletion} & \rotatebox{90}{Hover information} & \rotatebox{90}{Jump-to-definition} & \rotatebox{90}{Refactoring} & \rotatebox{90}{Live diagnostics} & \rotatebox{90}{Animation} & \rotatebox{90}{Proof visualizer} \\ \hline
VSCode-PVS    &  1.0.12 & VSCode & $\circletfill$  & $\circletfill$ & $\circletfill$  & $\circlet$ &  $\circletfill$  & $\circletfill$ & $\circletfill$ \\ \hline
PVS-Emacs       & 6.0 & Emacs & $\circletfillhb$ & $\circletcross$ & $\circletfillhb$  & $\circlet$ &  $\circletcross$  & $\circletfill$  & $\circletfill$   \\ \hline
CoqIDE              & 8.9.0 & N/A & $\circletcross$ & $\circletcross$ & $\circletcross$ & $\circletcross$  & $\circletcross$ & $\circletcross$ &  $\circletfillhb$   \\ \hline
SublimeHOL      & 2018 & Sublime &  $\circletfillhb$ & $\circletcross$ & $\circletcross$ & $\circletcross$  & $\circletcross$ & $\circletcross$ &  $\circletfillhb$   \\ \hline
Proof General    & 4.5 & Emacs, Eclipse & $\circletfillhb$ & $\circletcross$ & $\circletcross$ & $\circletcross$ & $\circletcross$ & $\circletcross$ & $\circletfillhb$   \\ \hline
Isabelle/jEdit      & 2019 & jEdit & $\circletfill$  & $\circletfill$  & $\circletfill$  & $\circlet$ & $\circletfill$  & $\circletfillhb$ &  $\circletfill$ \\ \hline
Lean                  & 0.14.1 & VSCode & $\circletfill$    & $\circletfill$  & $\circletfill$  & $\circletcross$  & $\circletfill$ & $\circletcross$  &  $\circletfill$  \\ \hline
IntelliJ-Arend       & 1.0.0 & IntelliJ IDEA & $\circletfill$    & $\circletfill$  & $\circletfill$  & $\circletfill$  & $\circletfill$ & $\circletcross$  & $\circletcross$  \\ \hline
KeYmaeraX       & 4.6.3 & N/A & $\circletcross$    & $\circletcross$  & $\circletcross$  & $\circletcross$  & $\circletcross$ & $\circletcross$  & $\circletfill$  \\ \hline
\end{tabularx}
\caption{Overview of the comparison results. The following symbols summarize the characteristics of a feature: fully implemented ($\protect\circletfill$); basic implementation ($\protect\circletfillhb$); planned feature ($\protect\circlet$); not available ($\protect\circletcross$).}\label{tab:results}
\end{table}

\section{Related Work}
\label{sec:related}
VSCode-PVS aims to align the functionalities of the PVS front-end to those of program analyzers such as Dafny~\cite{leino2010dafny}, or reasoning engines like Imandra~\cite{passmore2017formal}.
The front-end of these tools provides all functionalities typically available in modern IDEs for programming languages, including context-aware help and an integrated debugger.
Verification is carried out in the background, by continuously querying a pool of solvers while the user types the code.
Design solutions are adopted to keep the interface responsive and provide an overall smooth programming experience to the user.

The verification technology used by VSCode-PVS is not automatic, as in the case of Dafny and Imandra.
However, when the PVS analysis targets routine tasks such as discharging proof obligations necessary to prove type correctness, 
automatic analysis is usually feasible thanks to the powerful proof strategies provided by PVS. This opportunity needs to be exploited, as it would allow the completion of simple but time-consuming activities that developers need to carry out while creating a formal specification. Appropriate mechanisms need to be developed to limit the use of CPU time and memory resources that could make the interface not responsive. The split architecture adopted in VSCode-PVS and the asynchronous nature of the LSP protocol facilitate the implementation of these mechanisms. The possibility of creating an integrated debugger for executable fragments of a PVS specification is also another interesting option that needs to be explored for VSCode-PVS.
The Visual Studio Code editor provides already the graphical elements necessary for interacting with the logic of the debugger, including break-points, an interactive panel with the usual run/step-into/step-over commands, as well as an interactive view for inspecting the value of variables, call stack, etc. These elements need to be customized for the PVS language, and appropriate hooks need to be implemented in the back-end to provide the logic necessary for debugging.

In~\cite{pit2016company}, a Proof General plugin is developed that introduces syntax highlighting and autocompletion for the Coq specification language.
In~\cite{RabeUITP2014}, a generic user interface for theorem proving systems is introduced. The editor front-end builds on jEdit, and a prototypical specification language for declaring formal terms such as theories, terms, and context. This approach proves useful to implementing a generic version of basic front-end features such as autocompletion, abstract syntax display, error highlighting, and tooltips.
These and other similar efforts are certainly worth exploring. However, it is unclear if in the long run they will stand against the rapid evolution of editors such as Visual Studio Code and Atom.
%All main theorem proving systems and new reasoning engines are exploring these options. Examples include: Lean\footnote{\url{https://github.com/leanprover/vscode-lean}}, Isabelle\footnote{\url{https://github.com/seL4/isabelle/tree/master/src/Tools/VSCode}}; Coq\footnote{\url{https://github.com/siegebell/vscoq}}, Imandra\footnote{\url{https://github.com/AestheticIntegration/imandra-vscode}}.

\section{Conclusion and Future Directions}
\label{sec:conclu}
A new development environment for the PVS verification system has been presented that aims to align the PVS front-end to that of main stream tools used by software developers.
A split architecture is adopted, where an editor front-end communicates with a server back-end.
The back-end uses process workers to adapt the APIs of PVS to the Language Server Protocol, a de-facto standard communication protocol for code editors and analyzers.
The editor front-end builds on the features of Visual Studio Code, a modern open-source code editor.

VSCode-PVS is under active development. The environment is still in its infancy, but it already advances the standard Emacs front-end of PVS in many respects --- live diagnostics, context-sensitive auto-completion, point-and-click navigation, interactive tree-based view for proof exploration.

Previous attempts carried out by others to develop a new front-end for PVS had little success. One attempt aimed to integrate PVS in Eclipse. Difficulties were encountered to align the APIs provided by PVS to the functionalities required for Eclipse, and the development was ultimately abandoned. Another attempt involved the development of a Python front-end for PVS, using the wxPython\footnote{\url{https://wxpython.org}} graphic library. A simple interface was created to exchange commands with PVS. Fragments of these implementations can be found in the GitHub repository of PVS\footnote{\url{https://github.com/SRI-CSL/PVS}}.

Current work on the VSCode-PVS front-end focuses on the integration with the next release of PVS, which provides a new XMLRPC interface that will improve performance and robustness of the language server.
The creation of an integrated debugger is also planned. It will align the functionalities of the PVSio evaluator to those of debuggers used in programming languages.
Integration with the PVSio-web~\cite{masci2015pvsio} prototyping environment is another future direction. PVSio-web enables the creation of interactive prototypes based on formal models. The prototypes resemble the visual appearance of a final system. They can be used to create scenario-based simulations that facilitate engagement between PVS experts and developers that are not familiar with PVS or formal methods (see~\cite{masci-afford19} for application examples and success stories).

\smallskip\noindent
{\small
{\bf Acknowledgement.} Work by the first author is supported by NASA's System Wide Safety Project under NASA/NIA Cooperative Agreement NNL09AA00A.
}

%\nocite{*}
\bibliography{vscode-pvs}

\begin{thebibliography}{10}
\providecommand{\bibitemdeclare}[2]{}
\providecommand{\surnamestart}{}
\providecommand{\surnameend}{}
\providecommand{\urlprefix}{Available at }
\providecommand{\url}[1]{\texttt{#1}}
\providecommand{\href}[2]{\texttt{#2}}
\providecommand{\urlalt}[2]{\href{#1}{#2}}
\providecommand{\doi}[1]{doi:\urlalt{http://dx.doi.org/#1}{#1}}
\providecommand{\bibinfo}[2]{#2}

\bibitemdeclare{inproceedings}{aspinall2000proof}
\bibitem{aspinall2000proof}
\bibinfo{author}{David \surnamestart Aspinall\surnameend}
  (\bibinfo{year}{2000}): \emph{\bibinfo{title}{Proof General: A generic tool
  for proof development}}.
\newblock In: {\sl \bibinfo{booktitle}{International Conference on Tools and
  Algorithms for the Construction and Analysis of Systems}},
  \bibinfo{organization}{Springer}, pp. \bibinfo{pages}{38--43},
  \doi{10.1007/3-540-46419-0_3}.

\bibitemdeclare{inproceedings}{fulton2015keymaera}
\bibitem{fulton2015keymaera}
\bibinfo{author}{Nathan \surnamestart Fulton\surnameend},
  \bibinfo{author}{Stefan \surnamestart Mitsch\surnameend},
  \bibinfo{author}{Jan-David \surnamestart Quesel\surnameend},
  \bibinfo{author}{Marcus \surnamestart V{\"o}lp\surnameend} \&
  \bibinfo{author}{Andr{\'e} \surnamestart Platzer\surnameend}
  (\bibinfo{year}{2015}): \emph{\bibinfo{title}{KeYmaera X: An axiomatic
  tactical theorem prover for hybrid systems}}.
\newblock In: {\sl \bibinfo{booktitle}{International Conference on Automated
  Deduction}}, \bibinfo{organization}{Springer}, pp. \bibinfo{pages}{527--538},
  \doi{10.1007/978-3-319-21401-6_36}.

\bibitemdeclare{inproceedings}{leino2010dafny}
\bibitem{leino2010dafny}
\bibinfo{author}{K~Rustan~M \surnamestart Leino\surnameend}
  (\bibinfo{year}{2010}): \emph{\bibinfo{title}{Dafny: An automatic program
  verifier for functional correctness}}.
\newblock In: {\sl \bibinfo{booktitle}{International Conference on Logic for
  Programming Artificial Intelligence and Reasoning}},
  \bibinfo{organization}{Springer}, pp. \bibinfo{pages}{348--370},
  \doi{10.1007/978-3-642-17511-4_20}.

\bibitemdeclare{inproceedings}{masci-afford19}
\bibitem{masci-afford19}
\bibinfo{author}{Paolo \surnamestart Masci\surnameend} (\bibinfo{year}{2019 (to
  appear)}): \emph{\bibinfo{title}{Experiences on Streamlining Formal Methods
  Tools}}.
\newblock In: {\sl \bibinfo{booktitle}{International Workshop on Practical
  Formal Verification for Software Dependability (AFFORD'19)}}.

\bibitemdeclare{inproceedings}{masci2015pvsio}
\bibitem{masci2015pvsio}
\bibinfo{author}{Paolo \surnamestart Masci\surnameend},
  \bibinfo{author}{Patrick \surnamestart Oladimeji\surnameend},
  \bibinfo{author}{Yi~\surnamestart Zhang\surnameend}, \bibinfo{author}{Paul
  \surnamestart Jones\surnameend}, \bibinfo{author}{Paul \surnamestart
  Curzon\surnameend} \& \bibinfo{author}{Harold \surnamestart
  Thimbleby\surnameend} (\bibinfo{year}{2015}): \emph{\bibinfo{title}{PVSio-web
  2.0: Joining PVS to HCI}}.
\newblock In: {\sl \bibinfo{booktitle}{International Conference on Computer
  Aided Verification}}, \bibinfo{organization}{Springer}, pp.
  \bibinfo{pages}{470--478}, \doi{10.1007/978-3-319-21690-4_30}.

\bibitemdeclare{inproceedings}{masci2014formal}
\bibitem{masci2014formal}
\bibinfo{author}{Paolo \surnamestart Masci\surnameend},
  \bibinfo{author}{Yi~\surnamestart Zhang\surnameend}, \bibinfo{author}{Paul
  \surnamestart Jones\surnameend}, \bibinfo{author}{Paul \surnamestart
  Curzon\surnameend} \& \bibinfo{author}{Harold \surnamestart
  Thimbleby\surnameend} (\bibinfo{year}{2014}): \emph{\bibinfo{title}{Formal
  verification of medical device user interfaces using PVS}}.
\newblock In: {\sl \bibinfo{booktitle}{International Conference on Fundamental
  Approaches to Software Engineering}}, \bibinfo{organization}{Springer}, pp.
  \bibinfo{pages}{200--214}, \doi{10.1007/978-3-642-54804-8_14}.

\bibitemdeclare{inproceedings}{de2015lean}
\bibitem{de2015lean}
\bibinfo{author}{Leonardo \surnamestart de~Moura\surnameend},
  \bibinfo{author}{Soonho \surnamestart Kong\surnameend},
  \bibinfo{author}{Jeremy \surnamestart Avigad\surnameend},
  \bibinfo{author}{Floris \surnamestart Van~Doorn\surnameend} \&
  \bibinfo{author}{Jakob \surnamestart von Raumer\surnameend}
  (\bibinfo{year}{2015}): \emph{\bibinfo{title}{The Lean theorem prover (System
  Description)}}.
\newblock In: {\sl \bibinfo{booktitle}{International Conference on Automated
  Deduction}}, \bibinfo{organization}{Springer}, pp. \bibinfo{pages}{378--388},
  \doi{10.1007/978-3-319-21401-6_26}.

\bibitemdeclare{techreport}{munoz2003rapid}
\bibitem{munoz2003rapid}
\bibinfo{author}{C{\'e}sar~A \surnamestart Mu{\~n}oz\surnameend}
  (\bibinfo{year}{2003}): \emph{\bibinfo{title}{Rapid prototyping in PVS}}.
\newblock \bibinfo{type}{Technical Report},
  \bibinfo{institution}{{{NASA/CR-2003-212418, NIA Report No. 2003-03}}}.
\newblock
  \urlprefix\url{https://ntrs.nasa.gov/archive/nasa/casi.ntrs.nasa.gov/20040046914.pdf}.

\bibitemdeclare{inproceedings}{munoz2012}
\bibitem{munoz2012}
\bibinfo{author}{C{\'e}sar~A. \surnamestart Mu{\~{n}}oz\surnameend} \&
  \bibinfo{author}{Ramiro~A. \surnamestart Demasi\surnameend}
  (\bibinfo{year}{2012}): \emph{\bibinfo{title}{Advanced Theorem Proving
  Techniques in PVS and Applications}}.
\newblock In \bibinfo{editor}{Bertrand \surnamestart Meyer\surnameend} \&
  \bibinfo{editor}{Martin \surnamestart Nordio\surnameend}, editors: {\sl
  \bibinfo{booktitle}{Tools for Practical Software Verification: LASER,
  International Summer School 2011, Elba Island, Italy, Revised Tutorial
  Lectures}}, \bibinfo{publisher}{Springer Berlin Heidelberg},
  \bibinfo{address}{Berlin, Heidelberg}, pp. \bibinfo{pages}{96--132},
  \doi{10.1007/978-3-642-35746-6_4}.

\bibitemdeclare{inproceedings}{owre1992pvs}
\bibitem{owre1992pvs}
\bibinfo{author}{Sam \surnamestart Owre\surnameend}, \bibinfo{author}{John~M
  \surnamestart Rushby\surnameend} \& \bibinfo{author}{Natarajan \surnamestart
  Shankar\surnameend} (\bibinfo{year}{1992}): \emph{\bibinfo{title}{{PVS: A
  Prototype Verification System}}}.
\newblock In: {\sl \bibinfo{booktitle}{International Conference on Automated
  Deduction}}, \bibinfo{organization}{Springer}, pp. \bibinfo{pages}{748--752},
  \doi{10.1007/3-540-55602-8_217}.

\bibitemdeclare{techreport}{pvs-system-guide}
\bibitem{pvs-system-guide}
\bibinfo{author}{Sam \surnamestart Owre\surnameend}, \bibinfo{author}{Natarajan
  \surnamestart Shankar\surnameend}, \bibinfo{author}{John~M \surnamestart
  Rushby\surnameend} \& \bibinfo{author}{David~WJ \surnamestart
  Stringer-Calvert\surnameend} (\bibinfo{year}{1999}):
  \emph{\bibinfo{title}{PVS system guide}}.
\newblock \bibinfo{type}{Technical Report}, \bibinfo{institution}{Computer
  Science Laboratory, SRI International, Menlo Park, CA}.
\newblock \urlprefix\url{https://pvs.csl.sri.com/doc/pvs-system-guide.pdf}.

\bibitemdeclare{article}{sosym19}
\bibitem{sosym19}
\bibinfo{author}{M.~\surnamestart Palmieri\surnameend},
  \bibinfo{author}{C.~\surnamestart Bernardeschi\surnameend} \&
  \bibinfo{author}{P.~\surnamestart Masci\surnameend} (\bibinfo{year}{2019}):
  \emph{\bibinfo{title}{A Framework for FMI-based Co-Simulation of
  Human-Machine Interfaces}}.
\newblock {\sl \bibinfo{journal}{Software and Systems Modeling}},
  \doi{10.1007/s10270-019-00754-9}.

\bibitemdeclare{inproceedings}{palmieri2017co}
\bibitem{palmieri2017co}
\bibinfo{author}{Maurizio \surnamestart Palmieri\surnameend},
  \bibinfo{author}{Cinzia \surnamestart Bernardeschi\surnameend} \&
  \bibinfo{author}{Paolo \surnamestart Masci\surnameend}
  (\bibinfo{year}{2018}): \emph{\bibinfo{title}{Co-simulation of
  Semi-autonomous Systems: The Line Follower Robot Case Study}}.
\newblock In \bibinfo{editor}{Antonio \surnamestart Cerone\surnameend} \&
  \bibinfo{editor}{Marco \surnamestart Roveri\surnameend}, editors: {\sl
  \bibinfo{booktitle}{Software Engineering and Formal Methods}},
  \bibinfo{publisher}{Springer International Publishing}, pp.
  \bibinfo{pages}{423--437}, \doi{10.1007/978-3-319-74781-1_29}.

\bibitemdeclare{inproceedings}{passmore2017formal}
\bibitem{passmore2017formal}
\bibinfo{author}{Grant~Olney \surnamestart Passmore\surnameend} \&
  \bibinfo{author}{Denis \surnamestart Ignatovich\surnameend}
  (\bibinfo{year}{2017}): \emph{\bibinfo{title}{Formal verification of
  financial algorithms}}.
\newblock In: {\sl \bibinfo{booktitle}{International Conference on Automated
  Deduction}}, \bibinfo{organization}{Springer}, pp. \bibinfo{pages}{26--41},
  \doi{10.1007/978-3-319-63046-5_3}.

\bibitemdeclare{inproceedings}{pit2016company}
\bibitem{pit2016company}
\bibinfo{author}{Cl\'ement \surnamestart Pit-Claudel\surnameend} \&
  \bibinfo{author}{Pierre \surnamestart Courtieu\surnameend}
  (\bibinfo{year}{2016}): \emph{\bibinfo{title}{{Company-Coq: Taking Proof
  General one step closer to a real IDE}}}.
\newblock In: {\sl \bibinfo{booktitle}{{CoqPL'16: The Second International
  Workshop on Coq for PL}}}, \bibinfo{publisher}{Zenodo},
  \doi{10.5281/zenodo.44331}.

\bibitemdeclare{inproceedings}{platzer2008keymaera}
\bibitem{platzer2008keymaera}
\bibinfo{author}{Andr{\'e} \surnamestart Platzer\surnameend} \&
  \bibinfo{author}{Jan-David \surnamestart Quesel\surnameend}
  (\bibinfo{year}{2008}): \emph{\bibinfo{title}{KeYmaera: A hybrid theorem
  prover for hybrid systems}}.
\newblock In: {\sl \bibinfo{booktitle}{International Joint Conference on
  Automated Reasoning}}, \bibinfo{organization}{Springer}, pp.
  \bibinfo{pages}{171--178}, \doi{10.1007/978-3-540-71070-7_15}.

\bibitemdeclare{article}{RabeUITP2014}
\bibitem{RabeUITP2014}
\bibinfo{author}{Florian \surnamestart Rabe\surnameend} (\bibinfo{year}{2014}):
  \emph{\bibinfo{title}{{A Logic-Independent IDE}}}.
\newblock {\sl \bibinfo{journal}{Electronic Proceedings in Theoretical Computer
  Science}} \bibinfo{volume}{167}, p. \bibinfo{pages}{48–60},
  \doi{10.4204/eptcs.167.7}.

\bibitemdeclare{inproceedings}{wenzel2012isabelle}
\bibitem{wenzel2012isabelle}
\bibinfo{author}{Makarius \surnamestart Wenzel\surnameend}
  (\bibinfo{year}{2012}): \emph{\bibinfo{title}{Isabelle/jEdit: A Prover IDE
  within the PIDE framework}}.
\newblock In: {\sl \bibinfo{booktitle}{International Conference on Intelligent
  Computer Mathematics}}, \bibinfo{organization}{Springer}, pp.
  \bibinfo{pages}{468--471}, \doi{10.1007/978-3-642-31374-5_38}.

\bibitemdeclare{inproceedings}{wenzel2018isabelle}
\bibitem{wenzel2018isabelle}
\bibinfo{author}{Makarius \surnamestart Wenzel\surnameend}
  (\bibinfo{year}{2018}): \emph{\bibinfo{title}{Isabelle/PIDE after 10 years of
  development}}.
\newblock In: {\sl \bibinfo{booktitle}{UITP workshop: User Interfaces for
  Theorem Provers}}.
\newblock \urlprefix\url{https://sketis.
  net/wp-content/uploads/2018/08/isabellepide-uitp2018.pdf}.

\end{thebibliography}
\bibliographystyle{eptcs}
\end{document}